\documentclass{ws-ijmpe}
\usepackage[super,compress]{cite}
\usepackage{hyperref}       
\usepackage{amsmath}
\usepackage{subfig}
\usepackage{adjustbox}
\usepackage{graphicx}
\usepackage{booktabs}       
\usepackage{amsfonts}       
\usepackage{nicefrac}       
\usepackage{microtype}
\usepackage{caption}
\begin{document}


\catchline{}{}{}{}{}

\title{Constraining Bag constant for Hybrid Neutron stars}

\author{Ishfaq A. Rather\footnote{ishfaqrather81@gmail.com} $^{,\mathsection}$, Ankit Kumar$^{\dagger,\ddagger}$, Harish Chandra Das$^{\dagger,\ddagger}$, M. Imran$^\mathsection$, A. A. Usmani$^\mathsection$ and S. K. Patra $^{\dagger,\ddagger}$}

\address{$^\mathsection$ Department of Physics, Aligarh Muslim University,\\
	 Aligarh 202002, India}


\address{$^\dagger$ Institute of Physics, Bhubaneswar 751005, India\\
$^\ddagger$ Homi Bhabha National Institute, Training School Complex,\\ Anushakti Nagar, Mumbai 400085, India}


\maketitle


\begin{abstract}
We study the star matter properties for Hybrid equation of state (EoS) by varying the bag constant. We use the Effective-Field-Theory motivated Relativistic Mean-Field model (E-RMF) for hadron phase with recently reported FSUGarnet, G3 and IOPB-I parameter sets. The results of NL3 and NL3${\omega \rho}$ sets are also shown for comparison. The simple MIT Bag model is applied for the quark phase to construct the hybrid EoS. The hybrid neutron star mass and radius are calculated by varying with  $B^{1/4}$ to constrain the $B^{1/4}$ values. It is found that  $B^{1/4}$=130-160 MeV is suitable for explaining the quark matter in neutron stars.
\end{abstract}



\section{Introduction}
The existence of quark matter in the interior of hot dense objects like neutron stars was first pointed out by Witten \cite{PhysRevD.30.272,PhysRevD.30.2379} . This assumption of quark matter being completely stable may turn out to be the true ground state of hadronic matter. This quark matter can exist either as a pure phase in the central regions or a mixed-phase with hadronic matter \cite{PhysRevD.46.1274} . The presence of quarks inside the neutron stars will change the star matter properties.\\
 
 In the outer part of the neutron star (NS), where the density is low, the equation of state (EoS) is described by hadronic matter. With the increasing density, there is a possible phase transition from hadron matter to quark matter. The hadron-quark mixed phase is formed over a density range, thereafter a pure quark phase exists.  The density range over which a mixed-phase exists is determined by the charge neutrality condition. The information about the EoS of such stars, called Hybrid stars, can be studied and analyzed using the measurements from the recent gravitational wave constraints and the massive NS (PSR J0740+6620) M=$2.14^{+0.20}_{-0.18}$ $M_{\odot}$ \cite{Cromartie2020} . \\
 
 To study hybrid star EoS, we employ Effective-field-theory motivated Relativistic Mean-Field (E-RMF) model\cite{FURNSTAHL1996539} for the hadronic matter. This model has been very successful in determining the properties of finite \cite{HOROWITZ1981503,BOGUTA1977413,GAMBHIR1990132,RING1996193} as well as an infinite nuclear matter \cite{Walecka:1974qa,Arumugam:2004ys} . For the quark phase, we use the simple  MIT bag model \cite{PhysRevD.9.3471,PhysRevD.17.1109,PhysRevD.30.2379} to describe the quark matter in hybrid stars. In this model, the quarks are assumed to be confined in a bag of finite dimensions. The quarks are assumed to have very low mass inside the bag as compared to that outside, where the mass is very high. To balance the behavior of the bag and to find its size, a bag constant $B$ is introduced as a constant energy density in the system. At the surface of the bag, the outward pressure produced by the quarks is balanced by the inward pressure $B$. Thus the quark pressure decreases with the increasing value of $B$ thereby influencing the structure of the star. With a very low mass of $u$ and $d$ quarks, the value of $B$ depends on the mass of the strange quark. The value of $B$ varies from $B^{1/4} \approx$ 145-160 MeV for massless strange quark \cite{Stergioulas2003} . This range narrows down with the increase in the mass of strange quark. However, different bag values have been used in the literature. The bag value  $B^{1/4} \approx$ 200 MeV is used in the QCD calculations by H. Satz \cite{Satz} . Also, following the  CERN-SPS and RHIC data, the bag constants are allowed to have a wider range \cite{PhysRevC.66.025802, refId0} . So, we can consider the bag constant as an effective free parameter.  \\
 
 The complete EoS that describes the neutron star from its outer crust to the inner core is usually built from three different parts; the outer crust EoS, inner crust EoS and the core EoS. For the density range $\approx$ 10$^{4}$-10$^{11}$ g/cm$^{-3}$, the outer crust is defined by Baym-Pethick-Sutherland (BPS) EOS \cite{bps}. The inner crust EoS that defines the neutron star in the density range $\approx$ 10$^{11}$-10$^{14}$ g/cm$^{-3}$ is chosen according to the model considered so that it matches with the outer crust EoS at the neutron drip and the core EoS at crust-core transition. Either the inner crust EoS should be from the same model as the core EoS or at least the slope parameter of symmetry energy should match. The core is described within the RMF approach followed by a phase transition to quark matter. All these EoS's are matched accordingly so that the pressure is an increasing function of energy density. It has been observed \cite{PhysRevC.94.035804,PhysRevC.94.015808} that the star matter properties like mass and radius do not depend upon the outer crust EoS, but the presence of inner crust EoS is crucial for neutron star radius as the variation of more than 0.5km is observed in the radius  of a 1.4$M_{\odot}$ star.   \\
 
    The value of bag constant plays an important role in determining the structure and properties of hybrid stars. The EoS parameters like pressure, energy density and maximum mass of hybrid star are affected by varying the bag value. Although the standard value of $B$  in the MIT bag model is taken as $B^{1/4} \approx$ 140 MeV \cite{PhysRevD.22.1198, PhysRevD.12.2060} , a wide range of values varying from small to large have been used \cite{Baym:2017whm,STEINER2000239,BUBALLA2005205,NOVIKOV1981301} . But the definite range of bag values for hybrid stars is yet to be obtained. It is important to obtain a definite range of bag values for hybrid stars that will correspond to its stable configuration. The proper choice of bag constant can explain the hybrid stars with constrains imposed from recent gravitational wave observation using the simple MIT bag model.\\
 
In the present study, we combine the hadron phase and the quark phase to build a single hybrid EoS. We determine the EoS for hybrid stars and  calculate the mass and radius for different EoS's to constrain the bag constant. \\
 
This paper is organized as follows: in section 2 we discuss the theoretical approaches employed to study the equation of state of different phases. For hadronic matter, we employ Effective-field-theory motivated Relativistic Mean-Field (E-RMF) model\cite{FURNSTAHL1996539} using recently proposed different parameter sets. In the Quark matter, we employ the MIT Bag Model to describe the quark matter in neutron stars . We then combine the two phases to form a mixed hadron-quark phase. In section 3, we discuss the nuclear matter (NM) properties for different parameter sets used. The EoS obtained for hybrid star is also discussed. In section 4, we calculate the star matter properties like mass and radius for the obtained hybrid EoSs by varying the bag constant. All the calculated results are discussed in this section. Finally the summary and conclusions are given in section 5.\\

\section{FORMALISM}
\label{sec:headings} 
The basic relativistic Lagrangian of the Effective Field Theory (EFT) for strong interaction \cite{Walecka:1974qa,Reinhard:1989zi,Serot_1992} at low energies contains the contribution from $\sigma, \omega$ and $\rho$ mesons without any self-coupling terms which is the original Walecka model \cite{HOROWITZ1981503} . The addition of self-coupling terms by Boguta and Bodmer minimized the value of determined parameters to an acceptable range.  Parameter sets like NL3 \cite{PhysRevC.55.540}, FSUGarnet \cite{CHEN2015284} and IOPB-I \cite{PhysRevC.97.045806}  are developed that produce the results well withing the range.  Further, the contribution from $\delta$ meson explained certain properties of high dense matter \cite{KUBIS1997191,PhysRevC.89.044001} .  The parameter set G3 \cite{KUMAR2017197} contains the contribution from delta meson also in addition to other self and cross-coupling terms. Although the $\delta$ meson contribution to finite nuclei is nominal, its contribution to high dense matter is large. Physical observables like symmetry energy, effective mass of nucleons, mass and radius of the neutron star are influenced by delta meson inculsion and hence its contribution in the model should be considered. 
Since, the inculsion of cross-couplings has a large impact on neutron-skin thickness, symmetry energy and radius of the neutron star, a systematic formalism based on naturalness and Naive Dimensional Analysis (NDA), the effective field theory motivated relativistic mean field (E-RMF) lagrangian is constructed. \\

The E-RMF Lagrangian with $\delta$ meson inculsion, exchange mesons ($\sigma,\omega$ and $\rho$) and all other self- and cross-couplings is given by ref. \cite{FURNSTAHL1996539,Kumara:2017bti,PhysRevC.97.045806} .
\begin{equation}\label{eq1}
\begin{split}
\mathcal{E} (r)  =\sum_{\alpha} \phi^{\dagger}_{\alpha} (r) \Biggl\{-i \alpha .\nabla +\beta [M-\Phi(r) -\tau_3 D(r)]+W(r) +\frac{1}{2}\tau_3 R(r)+\frac{1+\tau_3}{2} A(r)\\
-\frac{i\beta \alpha}{2M}.\Bigg(f_{\omega}\nabla W(r)+\frac{1}{2}f_{\rho}\tau_3\nabla R(r)\Bigg)\Biggr\} \phi_{\alpha}(r)
+\Bigg(\frac{1}{2}+\frac{k_3}{3!}\frac{\Phi(r)}{M}+\frac{k_4}{4!}\frac{\Phi^2(r)}{M^2}\Bigg) \frac{m_s^2}{g_s^2}\Phi^2 (r)\\
-\frac{\zeta_0}{4!}\frac{1}{g_{\omega}^2}W^4(r)+\frac{1}{2 g_s^2}\Bigg(1+\alpha_1 \frac{\Phi(r)}{M}\Bigg)(\nabla \Phi(r))^2 -\frac{1}{2 g_{\omega}^2}\Bigg(1+\alpha_2 \frac{\Phi(r)}{M}\Bigg) \times (\nabla W(r))^2\\
-\frac{1}{2} \Bigg(1+\eta_1\frac{\Phi(r)}{M}+\frac{\eta_2}{2}\frac{\Phi^2(r)}{M^2}\Bigg)\frac{m_{\omega}^2}{g_{\omega}^2} W^2(r)-\frac{1}{2 e^2}(\nabla A(r))^2 -\frac{1}{g_{\rho}^2}(\nabla R(r))^2 \\
-\frac{1}{2}\Bigg(1+\eta_{\rho}\frac{\Phi(r)}{M}\Bigg)\frac{m_{\rho}^2}{g_{\rho}^2} R^2 (r) -\Lambda_{\omega}(R^2 (r) W^2(r))+\frac{1}{2 g_{\delta}^2}(\nabla D(r))^2+ \frac{1}{2}\frac{m_{\delta}^2}{g_{\delta}^2} (D(r)^2),
\end{split}
\end{equation}

Here, $\Phi$,$W$,$R$,$D$, and $A$ are the $\sigma$,$\omega$,$\rho$,$\delta$, and the photon fields. $g_{\sigma}$, $g_{\omega}$, $g_{\rho}$,$g_{\delta}$ and $e^2/(4\pi)$ are the coupling constants and $m_{\sigma}$,$m_{\omega}$, $m_{\rho}$, and $m_{\delta}$ are their respective masses.  $k_3$ ($k_4$) and $\zeta_0$ are the self-interacting coupling constants of $\sigma$ and $\omega$ respectively, while as $\eta_1$, $\eta_2$, $\eta_{\rho}$ and $\Lambda_{\omega}$ are the non-linear cross-coupling constants. Using the variational principle, the field equations for baryons and mesons are solved. The Dirac equation corresponding to eq.\ref{eq1} becomes 
\begin{equation}
\frac{\partial}{\partial \phi^{\dagger}_{\alpha}(r)} \Big[\mathcal{E}(r)-\sum_{\alpha} \phi^{\dagger}_{\alpha}(r) \phi_{\alpha}(r)\Big]=0
\end{equation}
From the above equation, we obtains the mean-field equations for $\Phi$, $W$, $R$,$D$ and $A$.\par 
The energy density and pressure for the hadronic phase, obtained using the energy momentum tensor, are given as 

	\begin{equation} \label{eq3}
	\begin{split}
	\mathcal{E}_H= <0|T_{00}|0>=\frac{2}{(2\pi)^3}\int d^3k E^*_i (k)+\rho W+\frac{m_s^2 \Phi^2}{g_s^2}\Bigg(\frac{1}{2}+\frac{k_3}{3!}\frac{\Phi}{M}+\frac{k_4}{4!}\frac{\Phi^2}{M^2}\Bigg)\\-\frac{1}{4!}\frac{\zeta_0 W^4}{g_{\omega}^2}
	+\frac{1}{2}\rho_3 R -\frac{1}{2}m_{\omega^2}\frac{W^2}{g_{\omega}^2}\Bigg(1+\eta_1\frac{\Phi}{M}+\frac{\eta_2}{2}\frac{\Phi^2}{M^2}\Bigg)\\
	-\frac{1}{2}\Bigg(1+\frac{\eta_{\rho}\Phi}{M}\Bigg)\frac{m_{\rho}^2}{g_{\rho}^2}R^2
	 -\Lambda_{\omega}(R^2 W^2)
	+\frac{1}{2}\frac{m_{\delta}^2}{g_{\delta}^2}(D^2)+\sum_l \mathcal{E}_l,
	\end{split}
	\end{equation}
and\\
\\
	\begin{equation}\label{eq4}
	\begin{split}
	P_H= \frac{1}{3}\sum_{i=1}^{3} <0|T_{ii}|0> = \frac{2}{3(2\pi)^3}\int d^3k E^*_i (k)-\frac{m_s^2 \Phi^2}{g_s^2}\Bigg(\frac{1}{2}+\frac{k_3}{3!}\frac{\Phi}{M}+\frac{k_4}{4!}\frac{\Phi^2}{M^2}\Bigg)\\
	+\frac{1}{4!}\frac{\zeta_0 W^4}{g_{\omega}^2} +\frac{1}{2}m_{\omega^2}\frac{W^2}{g_{\omega}^2}\Bigg(1+\eta_1\frac{\Phi}{M}+\frac{\eta_2}{2}\frac{\Phi^2}{M^2}\Bigg)
	+\frac{1}{2}\Bigg(1+\frac{\eta_{\rho}\Phi}{M}\Bigg)\frac{m_{\rho}^2}{g_{\rho}^2}R^2\\
	 +\Lambda_{\omega}(R^2 W^2)
	-\frac{1}{2}\frac{m_{\delta}^2}{g_{\delta}^2}(D^2)+\sum_l P_l,
	\end{split}
	\end{equation}
where, \\
\\
$\mathcal{E}_l$ and $P_l$ are the energy density and pressure for leptons. $M_p^*$ and $M_n^*$ are the effective masses of proton and neutron.
For neutron star matter, the $\beta$-equilibrium and charge neutrality are two important conditions to be satisfied to determine the composition of the system. These are
\begin{equation}\label{c1}
\mu_p = \mu_n - \mu_e. 
\end{equation}
and
\begin{equation}
q_{total} = \sum_{i=n,p} q_i k_i^3/(3\pi^2)+\sum_l q_l k_l^3/(3\pi^2)=0,
\end{equation}

 For the quark phase, we employ MIT Bag model for the unpaired quark matter \cite{PhysRevD.9.3471,PhysRevD.17.1109,PhysRevD.30.2379} .
  It assumes the quarks to be confined in a large colorless region where the quarks are free to move. The quark masses are taken as $m_u$= $m_d$ =5.0 MeV and $m_s$ = 150 MeV. We ignore the one gluon exchange inside the gas. The equilibrium condition in the quark matter is given as
\begin{equation}\label{c2}
\mu_d=\mu_s=\mu_u + \mu_e.
\end{equation}
The chemical potential of the quarks follow from the two independent chemical potentials $\mu_n$ and $\mu_e$ as:
\begin{equation}
\mu_u =\frac{1}{3}\mu_n -\frac{2}{3}\mu_e,
\end{equation}
\begin{equation}
\mu_d =\frac{1}{3}\mu_n +\frac{1}{3}\mu_e,
\end{equation}
and
\begin{equation}
\mu_s =\frac{1}{3}\mu_n +\frac{1}{3}\mu_e.
\end{equation}
The charge neutrality condition is satisfied by
\begin{equation}
\frac{2}{3}n_u-\frac{1}{3}n_d -\frac{1}{3}n_s -n_e=0,
\end{equation}
where, $n_q (q=u,d,s,e)$ is the number density of quarks. The total quark density is
\begin{equation}
n_Q = \frac{1}{3}(n_u +n_d +n_s).
\end{equation}
The pressure of the quarks (q=u,d,s) is given by \cite{book}
\begin{equation}
P_Q = \frac{1}{4\pi^2}\sum_q \Biggl\{\mu_q k_q \Bigg(\mu_q^2 -\frac{5}{2}m_q^2\Bigg)+\frac{3}{2}m_q^4 ln \Bigg(\frac{\mu_q +k_q}{m_q}\Bigg)\Biggr\}.
\end{equation}
The total pressure due to quarks and leptons is given by
\begin{equation}\label{eq16}
P=P_Q +P_l -B,
\end{equation}
where, B is the Bag constant.\\

At high density, the boundary between the pure HM and QM is supposed to coexist, but it is not well defined \cite{PhysRevD.46.1274} . The density range over which a mixed-phase can exist is governed by the $\beta$-equilibrium and charge neutrality conditions. Various methods have been used in the literature concerning the quark-hadron phase transition in neutron stars \cite{PhysRevD.46.1274,PhysRevC.60.025801,PhysRevC.75.035808,PhysRevC.66.025802,PhysRevC.89.015806} . The Gibbs construction (GC) \cite{PhysRevD.46.1274} and the Maxwell construction (MC) \cite{PhysRevD.88.063001} are commonly used to describe the quark-hadron phase transition. It has been shown that beyond a certain limiting value of surface tension, the mixed-phase becomes mechanically unstable and hence Maxwell construction provides a more relevant way in describing the hybrid star properties \cite{PhysRevC.83.068801,PhysRevD.83.024014} . However, the surface tension at the mixed-phase is still not known properly and hence the formation of hadron-quark mixed phase by Gibbs construction is also possible even if the surface tension is too small. In the present work, we assume that the surface tension is not too high to ensure the formation of mixed phase by GC. \\

 The Gibbs conditions for the mixed phase are given as:\\
\begin{equation}
P_{HP}(\mu_{HP}) = P_{QP}(\mu_{QP}) = P_{MP},
\end{equation}
and
\begin{equation}
\mu_{HP,i} = \mu_{QP,i} = \mu_i, ~i=n,e.
\end{equation}
\vspace{0.5cm}
\section{Results and Discussions}
In this section, we discuss our results for hybrid EoS. For hadron phase we used different parameter sets NL3 \cite{PhysRevC.55.540} , FSUGarnet \cite{CHEN2015284} , G3 \cite{KUMAR2017197} and IOPB-I \cite{PhysRevC.97.045806} as shown in table \ref{tab:HM} . The NM  parameters for the EoS at saturation density $J$ , $L$, $K_{sym}$ and $Q_{sym}$ for all parameter sets are listed in Table \ref{tab:NM}. For NL3 set, the symmetry energy $J$ = 37.43 MeV and slope parameter $L$ = 118.65 MeV are little higher than the empirical value $J$ = 31.6 $\pm$ 2.66 MeV and $L$ = 58.9 $\pm$ 16 MeV  \cite{LI2013276} . The  $J$ and $L$ for other parameter sets lie well within the given range. The value of incompressibility $K$ for different parameter sets are compatible with the observational data from various experiments \cite{Colo:2013yta,Piekarewicz:2013bea} .  With the higher value of slope parameter for NL3 set, we have used NL3${\omega \rho}$ which includes the non-linear $\omega-\rho$ terms that softens the symmetry energy density dependence \cite{PhysRevLett.86.5647}.  NL3${\omega \rho}$ set has symmetry energy and its slope parameter as 31.66 and 55.21 MeV respectively which is much lower than the NL3 set \cite{PhysRevC.94.035804,PhysRevC.90.045803} . \\
\begin{table}[ht]
	\centering
	\caption{Parameter sets and the corresponding nuclear matter properties for hadron matter. For all the sets, the nucleon mass is $M$= 939.0 MeV. All the coupling constants are dimensionless except $k_3$ which has the dimensions of fm$^{-1}$. }
	\large
	\begin{tabular}{ ccccc }
		\midrule
		\midrule
		&NL3&FSUGarnet&G3&IOPB-1 \\
		\midrule
		$m_s/M$ & 0.541&0.529&0.559&0.533\\
		$m_{\omega}/M$ &0.833&0.833&0.832&0.833\\
		$m_{\rho}/M$&0.812&0.812&0.820&0.812\\
		$m_{\delta}/M$&0.0&0.0&1.043&0.0\\
		$g_s/{4\pi}$&0.813&0.837&0.782&0.827\\
		$g_{\omega}/{4\pi}$&1.024&1.091&0.923&1.062\\
		$g_{\rho}/{4\pi}$&0.712&1.105&0.962&0.885\\
		$g_{\delta}/{4\pi}$&0.0&0.0&0.160&0.0\\
		$k_3$&1.465&1.368&2.606&1.496\\
		$k_4$&-5.688&-1.397&1.694&-2.932\\
		$\zeta_0$&0.0&4.410&1.010&3.103\\
		$\eta_1$&0.0&0.0&0.424&0.0\\
		$\eta_2$&0.0&0.0&0.114&0.0\\
		$\eta_{\rho}$&0.0&0.0&0.645&0.0\\
		$\Lambda_{\omega}$&0.0&0.043&0.038&0.024\\
		$\alpha_1$&0.0&0.0&2.000&0.0\\
		$\alpha_2$&0.0&0.0&-1.468&0.0\\
		$f_{\omega}/4$&0.0&0.0&0.220&0.0\\
		$f_{\rho}/4$&0.0&0.0&1.239&0.0\\
		$\beta_{\sigma}$&0.0&0.0&-0.087&0.0\\
		$\beta_{\omega}$&0.0&0.0&-0.484&0.0\\
		\midrule
		\midrule
		\end{tabular}
	\label{tab:HM}
\end{table}
\begin{table}[ht]
	\centering
	\caption{Nuclear matter properties of hadron matter for different parameter sets. }
	\begin{tabular}{ cccccc }
		\midrule
		\midrule
		&NL3&NL3${\omega \rho}$&FSUGarnet&G3&IOPB-1 \\
		\midrule
		\midrule
		$\rho_0$ (fm$^{-3}$) & 0.148&0.148 &0.153&0.148&0.149\\
		$\epsilon_0 $(MeV) & -16.29&-16.30&-16.23&-16.02&-16.10 \\
		M*/M&0.595&0.602&0.578&0.699&0.593\\
		$J $(MeV)  & 37.43&31.66&30.95&31.84&33.30 \\
		$L $(MeV) &118.65&55.21&51.04&49.31&63.58 \\
		$K_{sym}$ (MeV) &101.34&-8.05&59.36&-106.07&-37.09 \\
		$Q_{sym}$ (MeV)& 177.90&188.80&130.93&915.47&862.70 \\
		$K$ (MeV)&271.38&272.04&229.5&243.96&222.65\\
		\midrule
		\midrule
	\end{tabular}
	\label{tab:NM}
\end{table}\\

 Fig. \ref{prho} shows the variation of pressure with the baryon density for pure neutron matter (PNM).  The results are compared with the experimental flow data obtained from the analysis of heavy ion collisions \cite{Danielewicz1592} ,where the upper one (stiff-expt.) corresponds to the strong density dependence of $S(\rho)$ and the lower one (soft-expt.) corresponds to the weak dependence. It is clear from Fig. \ref{prho} that the PNM EoS for G3 set is compatible with the experimental data. The NL3 set produces stiffer results than the other parameter sets at high densities. The IOPB-I and FSUGarnet EoS are also compatible with the data.\\
\begin{adjustbox}{center, caption={a) Pressure vs baryon density for Pure Neutron Matter (PNM) and b) EoS for NS matter in $\beta$-equilibrium and charge neutrality condition with different E-RMF parameters.},label={prho},nofloat=figure}
	\includegraphics[scale=0.25]{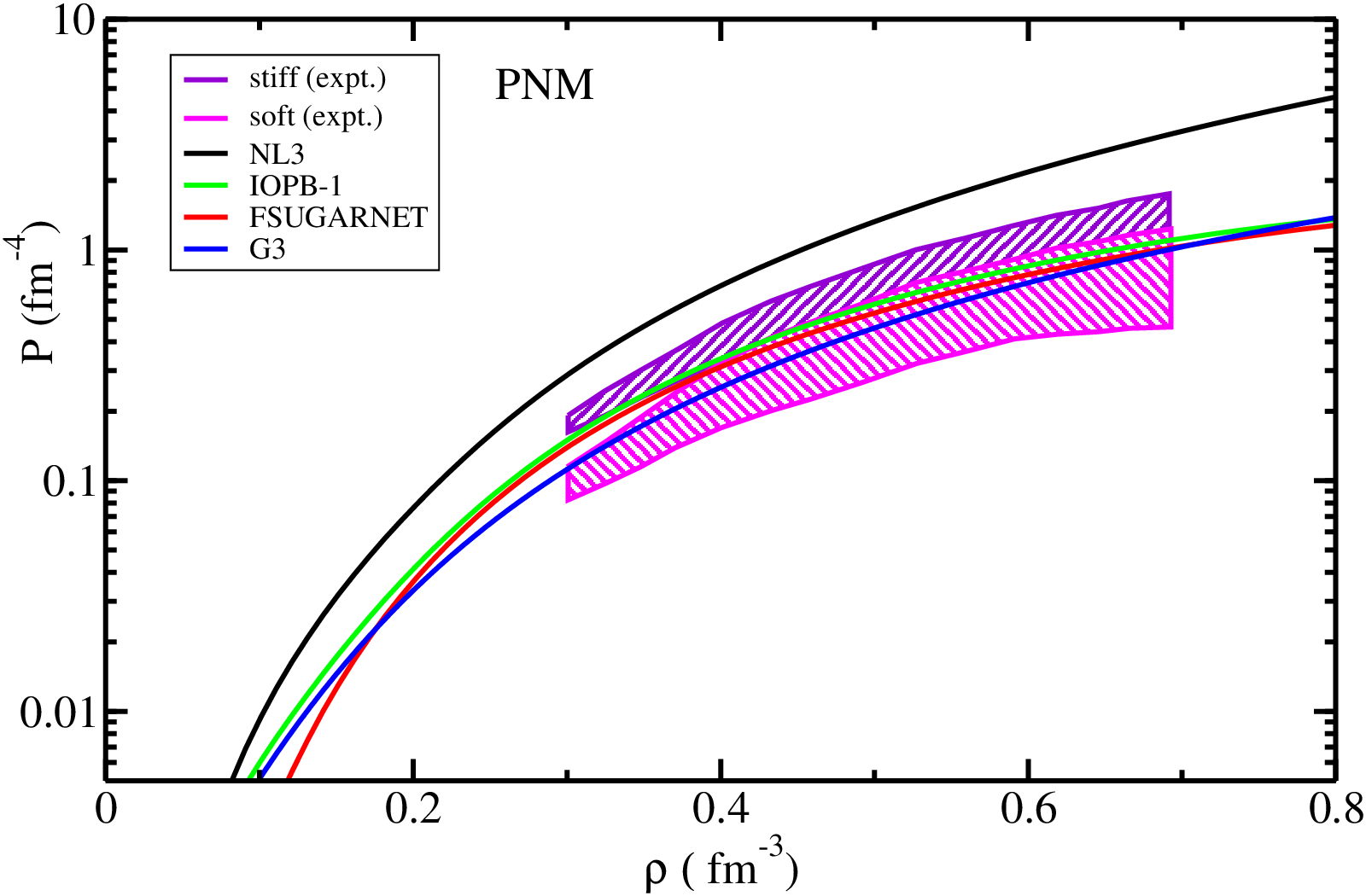}
	\includegraphics[scale=0.25]{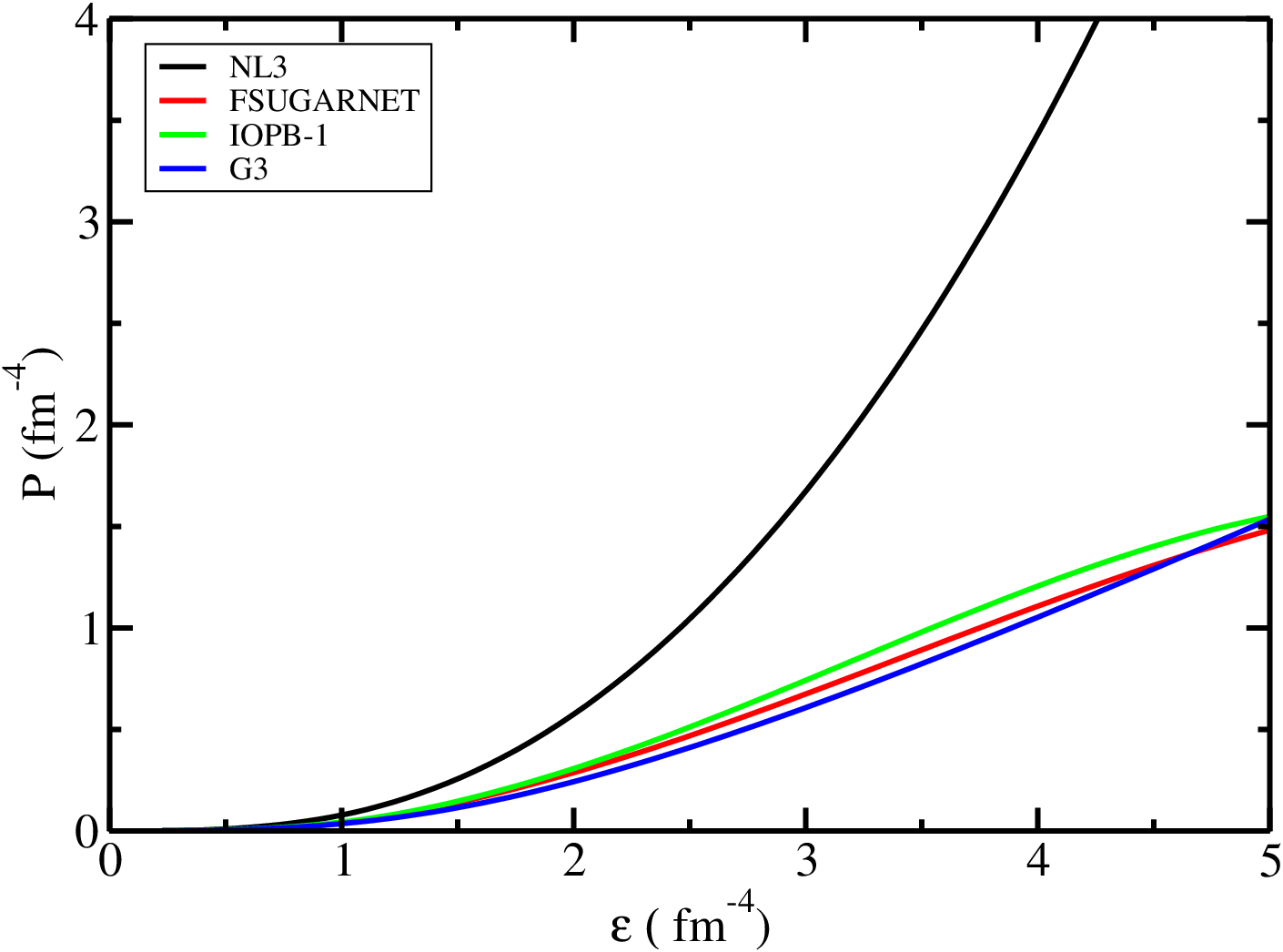}
\end{adjustbox} 
 The pressure vs energy density for $\beta$-equilibrated charge neutral neutron star matter  for parameter sets NL3, FSUGarnet, IOPB-I and G3 are also displayed.  The NL3 parameter set yields a stiffer EoS. FSUGarnet and IOPB-I  have similar EoSs at high density but they differ at low density. FSUGarnet has soft eos at low density but becomes stiff at higher density as compared to IOPB-I and G3. The G3 set provides the soft EoS than other parameter sets. This also shows that the addition of delta meson helps in obtaining soft EoS which results in the lower maximum mass and the radius of neutron star than others. FSUGarnet EoS is used only in the pure hadronic part as a comparison to other parameter sets. No calculation for the hybrid stars is done using the FSUGarnet parameter set.\par
The hybrid EoS for different hadronic matter parameter sets (NL3, IOPB-I and G3) and different quark matter bag values ($B^{1/4}$ = 100, 130, 160, 180 and 200 MeV) are shown in Fig.(\ref{eos_hybrid}).  The hybrid EoS for NL3${\omega \rho}$ set has also been plotted as a comparison with the NL3 set. The energy density increases with the bag values and hence the pressure will decrease. This means that the hybrid EoS becomes softer as the bag values increase.  The importance of hybrid EoS lies in the formation of a mixed-phase. The transition from HM to QM using Gibbs condition determines the stiffness or softness of the mixed-phase.\par
The transition density at which the hadron-quark phase transition takes place varies with the bag constant. For very small bag constants (100 MeV), the phase transition takes place at a density lower than the normal nuclear density ($\rho_0$=0.16$fm^{-3}$). This shows the presence of quarks at normal nuclear density level, which is unphysical because the quarks appear in the neutron stars at density a few times the normal nuclear density. For higher values, the hadron-quark mixed phase exists in the density range $\rho$= (2-7)$\rho_0$. It is found that both NL3 and NL3${\omega \rho}$ parameter sets show almost same phase transition properties at all bag constants. 
\begin{adjustbox}{center, caption={ Hybrid EoS for different bag constants for all parameter sets.},label={eos_hybrid},nofloat=figure}
	\includegraphics[scale=0.49]{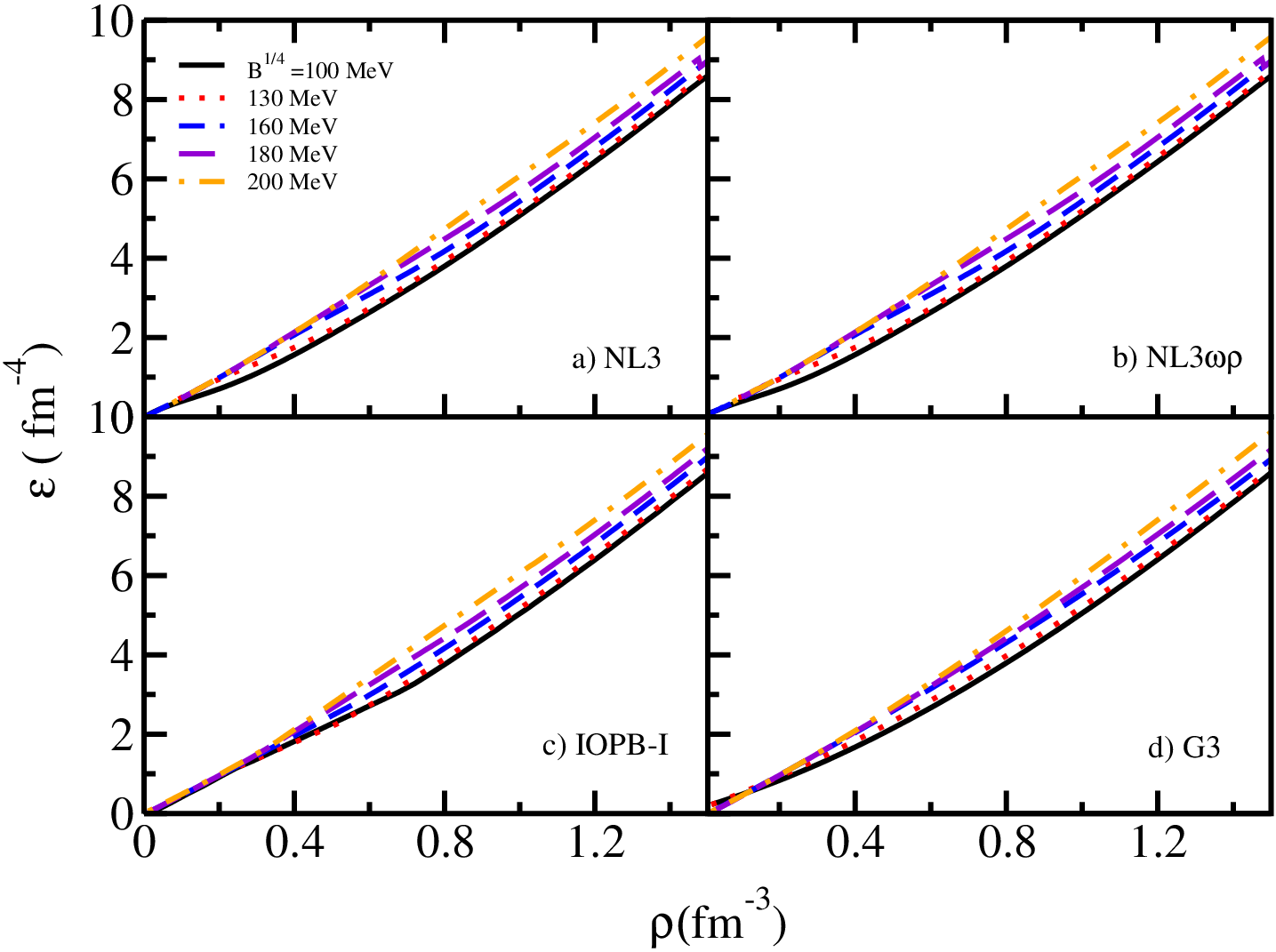}
\end{adjustbox}

\section{Structure and Properties of hybrid star}
 With the EoS obtained, we can calculate the properties of hybrid stars. Assuming the star to be spherical and stationary, the Tolman-Oppenheimer-Volkoff (TOV) equations \cite{PhysRev.55.374,PhysRev.55.364} are used to evaluate the structure of the star.
 \begin{equation}\label{tov1}
 \frac{dP(r)}{dr}= -G\frac{[\mathcal{E}(r) +P(r)][M(r)+4\pi r^3 P(r)]}{r^2(1-2M(r)/r) }
 \end{equation}
 and
 \begin{equation}\label{tov2}
 \frac{dM(r)}{dr}= 4\pi r^2 \mathcal{E}(r)
 \end{equation}
 where, $G$ is the gravitational constant and $M(r)$ is the gravitational mass. For a given EoS, the above equations can be solved for the given boundary conditions $P(0)=P_c$, $M(0)=0$, where $P_c$ is the central pressure. The value of radius ($r=R$) at which the pressure vanishes defines the surface of the star.\par 
 
 \vspace{1.0cm}
 \begin{adjustbox}{center, caption={ Mass Radius profile of pure hadronic matter for NL3, NL3${\omega \rho}$, IOPB-I, and G3 parameter sets. The recent constraints on mass and radius \cite{PhysRevLett.119.161101,PhysRevLett.120.172702} are also shown.},label={eos},nofloat=figure}
 	\includegraphics[scale=0.38]{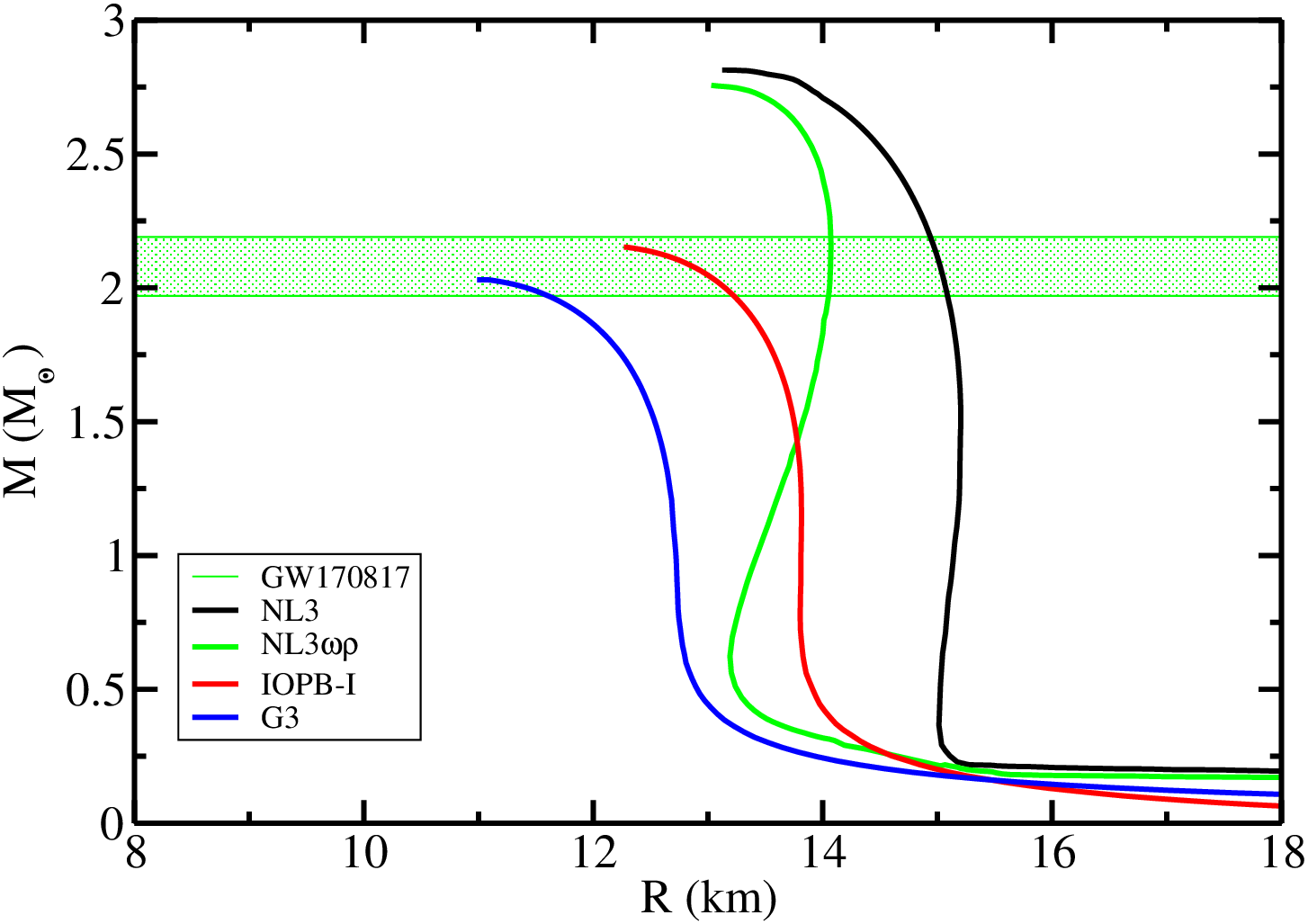}
 \end{adjustbox}
 Fig.\ref{eos} shows the mass radius profile obtained for pure hadronic matter using different parameter sets. The NL3 set predicts a large radius and mass for neutron star. The NS mass is found to be 2.81$M_{\odot}$ and the corresponding radius is 13.20 km. NL3${\omega \rho}$ produces a neutron star with a maximum mass of 2.75 M$_{\odot}$ which is quite close to NL3 NS maximum mass. This close resemblance is also seen in the NS radius. The NL3${\omega \rho}$ MR curve differs from the usual NL3 MR curve at low mass and low radii. This means that the  NL3${\omega \rho}$ set predicts a smaller radius of NS at the canonical mass. For the IOPB-I parameter set, the NS maximum mass is around 2.15$M_{\odot}$ and the radius is 12.27 km. The G3 set predicts an NS with a maximum mass of 2.03$M_{\odot}$ and the corresponding radius 11.06 km \cite{PhysRevC.97.045806} . The maximum mass of a non-rotating NS is in the range 2.01$\pm$0.04 $\le$ $M(M_{\odot})$ $\le$2.16$\pm$0.03. This range was obtained by Rezzolla $et$ $al$ \cite{Rezzolla_2018} by combining the recent gravitational wave observation of a binary neutron star merger (GW170817) \cite{PhysRevLett.119.161101} with the quasi-universal relation between rotating and non-rotating neutron star maximum mass.\\
 
 All the MR plots shown in fig.\ref{eos} represent the complete stellar EoS obtained by properly joining the inner crust EoS with outer crust and core EoS. The inner crust EoS for all the models is taken such that the symmetry energy properties of both the crust and core EoS match with each other. For IOPB-I and G3 families, we do not have a unified EoS for the inner crust. However considering the slope parameter $L$ of symmetry energy for these two families, we have used IU-FSU \cite{PhysRevC.82.055803}and FSU  \cite{PhysRevLett.95.122501}models as the inner crust EoS for G3 and IOPB-I respectively as they have a close comparison in slope parameter value. All the inner crust EoS's used can be found in ref. \cite{PhysRevC.90.045803}. From the plots, it is clear that an inner crust EoS with smaller value of slope parameter represents a large fraction of the total crust.\\

 \begin{figure}	
 	\centering
 	\includegraphics[width=13.5cm,height=11.5cm]{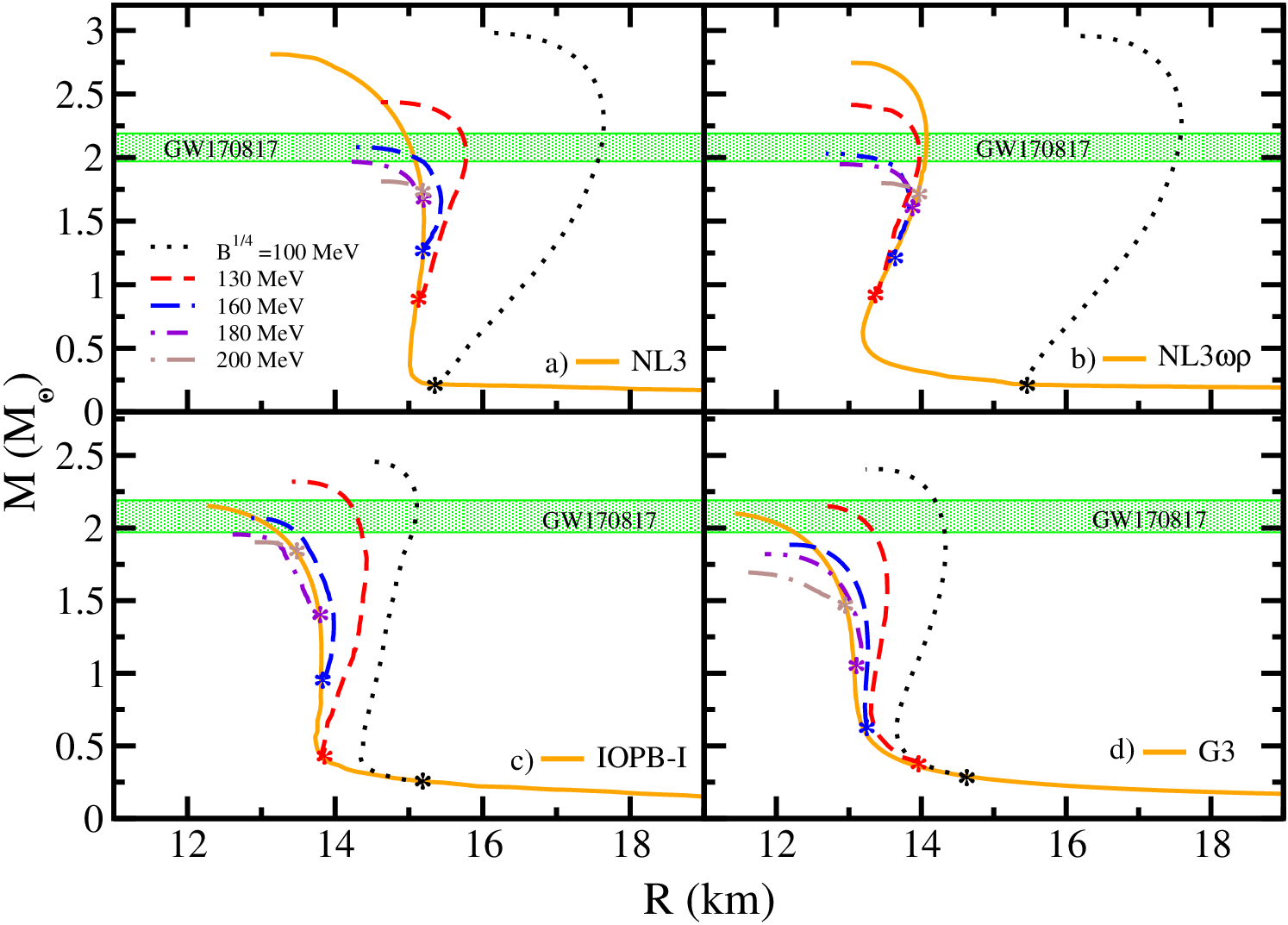}
 	\caption{Mass Radius profile of hybrid star for a) NL3, b) NL3${\omega \rho}$, c) IOPB-I and d) G3 parameter sets. The corresponding asterisks denote the transition from pure hadron star to hybrid star.}
 	\label{hyb}
 \end{figure}
The mass-radius profile for hybrid EoS obtained for different bag constants is shown in Fig.\ref{hyb}. The green band represents the maximum mass range obtained for a non-rotating star \cite{PhysRevLett.119.161101,Margalit_2017,Antoniadis1233232} . This band also satisfies the precisely measured mass of PSR J0348+0432 and PSR J1614-2230 with mass (2.01$\pm$0.04)$M_{\odot}$ \cite{Antoniadis1233232} and (1.97$\pm$0.04)$M_{\odot}$ \cite{Demorest} respectively. These measurements imply that the maximum mass of any NS predicted by  any theoretical model should reach the limit of $\approx$ 2.0 $M_{\odot}$. It is clear that the maximum mass of hybrid star decreases as the bag constant increases. For lower bag values, the maximum mass produced is very high at $\approx$ 2.5 $M_{\odot}$ for IOPB-I and G3 sets, and for higher values of bag, the mass is reduced to $\approx$ 1.8 $M_{\odot}$. The mass-radius profile of pure hadronic matter is also shown (solid lines ) again for all parameter sets for comparison. The mass of pure hadronic matter for IOPB-I and G3 sets lie well within the bag values $B^{1/4}$= 130-160 MeV. The MR curves for NL3 and NL3${\omega \rho}$ sets show a  similar trend for all bag constants. \\

\begin{table}[ht]
	\centering
	\caption{Phase transition properties of hybrid EoS for different bag constants. $\rho_{MP}$ represents the beginning of the formation of mixed phase. $P_{trans}$ represents the corresponding pressure in units of $10^{-1}$ fm$^{-4}$. $M_{Hys}$ and $R$ represent the mass and the corresponding radius where the pure hadron star transforms into a hybrid star. }
	\begin{tabular}{ cccccc }
		\midrule
		\midrule
		$B^{1/4}$ (MeV)&100&130&160&180&200 \\
		\midrule
		&&NL3\\
		\midrule
		$\rho_{MP} (\rho_0)$ &0.98&1.12&1.81&3.05&4.63\\
		$P_{trans}$	&0.42&0.56&1.58&3.64&5.46\\
		$M_{Hys}$($M_{\odot}$)  & 0.21&0.88&1.27&1.67&1.72 \\
		$R$ (km) & 15.37&15.09&15.19&15.24&15.19\\
		\midrule
		&&NL3${\omega \rho}$\\
	\midrule
	$\rho_{MP} (\rho_0)$ &0.97&1.10&1.79&3.01&4.58\\
	$P_{trans}$	&0.43&0.58&1.61&3.67&5.49\\
	$M_{Hys}$($M_{\odot}$)  & 0.21&0.88&1.23&1.62&1.72 \\
	$R$ (km) & 15.46&13.34&13.65&13.89&13.98\\
	\midrule
		&& IOPB-I\\
		\midrule	
			$\rho_{MP} (\rho_0)$ &0.96&1.04&1.63&2.96&4.42\\
		$P_{trans}$	&0.38&0.75&1.13&3.60&6.94\\
		$M_{Hys} $($M_{\odot}$)  & 0.24&0.42&0.96&1.39&1.83 \\
		$R$ (km) & 15.22&13.83&13.80&13.78&13.50\\
		\midrule
		&& G3\\
		\midrule	
		$\rho_{MP} (\rho_0)$ &0.96&1.02&1.58&2.84&4.16\\
	$P_{trans}$	&0.27&0.42&0.87&2.83&5.83\\
	$M_{Hys} $($M_{\odot}$)  & 0.28&0.35&0.61&1.05&1.46 \\
		$R$ (km) &14.62&13.94&13.24&13.07&12.95\\		
		
		\midrule
		\midrule
	\end{tabular}
	\label{tab:Trans}
\end{table}

  The asterisks denote the transition masses from hadron star to hybrid star. The transition to hybrid star begins earlier for low bag values. As the bag values increase, the phase transition also increases. The addition of inner crust EoS stabilizes the transition between hadronic matter and quark matter.   \\

Table \ref{tab:Trans} shows the properties of phase transition from the hadron phase to the mixed-phase. For low bag constants (100 MeV), the transition takes place at a density lower than the normal nuclear density, which is unphysical because it means that the quarks are present at nuclear matter density. For higher bag values, the transition takes place at densities around (2-5)$\rho_0$. $P_{trans}$ and $M_{Hys}$ represent respectively the corresponding transition pressure and the mass at which the pure hadron star makes a transition to hybrid star. The radius at which the transition takes place has also been shown in table \ref{tab:Trans}. Apart from bag value of 100 MeV, the hadron-quark transition takes place at an almost same radius of $\approx$ 15 km, 14 km, 14 km and 13 km for NL3, NL3${\omega \rho}$,IOPB-I, and G3 parameter sets respectively. \\

Despite the large difference in the symmetry energy slope parameter for NL3 and NL3${\omega \rho}$ sets, the hadron-quark phase transition properties for both parameter sets is very close except for the radius which differs by more than 1 km. Thus, a proper matching of crust EoS and core EoS is important to determine the true radius of a star. Also,  proper knowledge about the  inner crust EoS is important to properly calculate the star matter properties. \\

The mass at which the transition from pure hadron matter to hybrid star matter takes place increases with the bag constant, which means that as the bag constant increases, the hybrid star becomes more and more hadronic.\\
 
The maximum mass along with the corresponding radius of hybrid star matter for different parameter sets is shown in table \ref{mixp} for different bag constants. The calculations for the pure hadron matter are also shown. It s clear from the table that the maximum mass decreases with the bag constant. The radius also decreases except for NL3 and IOPB-I parameter sets at bag value of 200 MeV. \\

\vspace{2.5cm}
\begin{table}[ht]
	\centering
	\caption{Maximum mass and radius of hybrid stars for different bag constants. The results for pure hadron matter are also shown. }
	\begin{tabular}{ ccccccc }
		\midrule
		\midrule
		$B^{1/4}$ (MeV)&Pure Hadron&100&130&160&180&200 \\
		\midrule
		&&NL3\\
		\midrule	
		$M $($M_{\odot}$)  &2.81& 2.98&2.43&2.08&1.97&1.81 \\
		$R $(km) &13.20&16.15&14.72&14.39&14.13&14.65\\
	
		\midrule
		
		&&NL3${\omega \rho}$\\
		\midrule	
		$M $($M_{\odot}$)  &2.75& 2.95&2.41&2.03&1.94&1.79 \\
		$R $(km) &13.01&16.06&12.82&12.68&12.87&13.40\\
		
		\midrule
		&& IOPB-I\\
		\midrule	
		$M $($M_{\odot}$) &2.15 & 2.46&2.32&2.07&1.95&1.90 \\
		$R $(km) &12.27&14.45&13.50&12.91&12.64&13.00\\
		\midrule
		&& G3\\
		\midrule	
		$M $($M_{\odot}$) &2.03 & 2.40&2.15&1.88&1.82&1.69 \\
		$R $(km) &11.06&13.40&12.59&12.24&11.88&11.57\\
		\midrule
		\midrule
	\end{tabular}
	\label{mixp}
\end{table} 

\vspace{0.5cm}
\begin{adjustbox}{center, caption={ Variation of Maximum Mass and Radius of hybris stars with different bag constants for NL3, NL3${\omega \rho}$, IOPB-I and G3 parameter sets.},label={mrb},nofloat=figure}
	\includegraphics[scale=0.38]{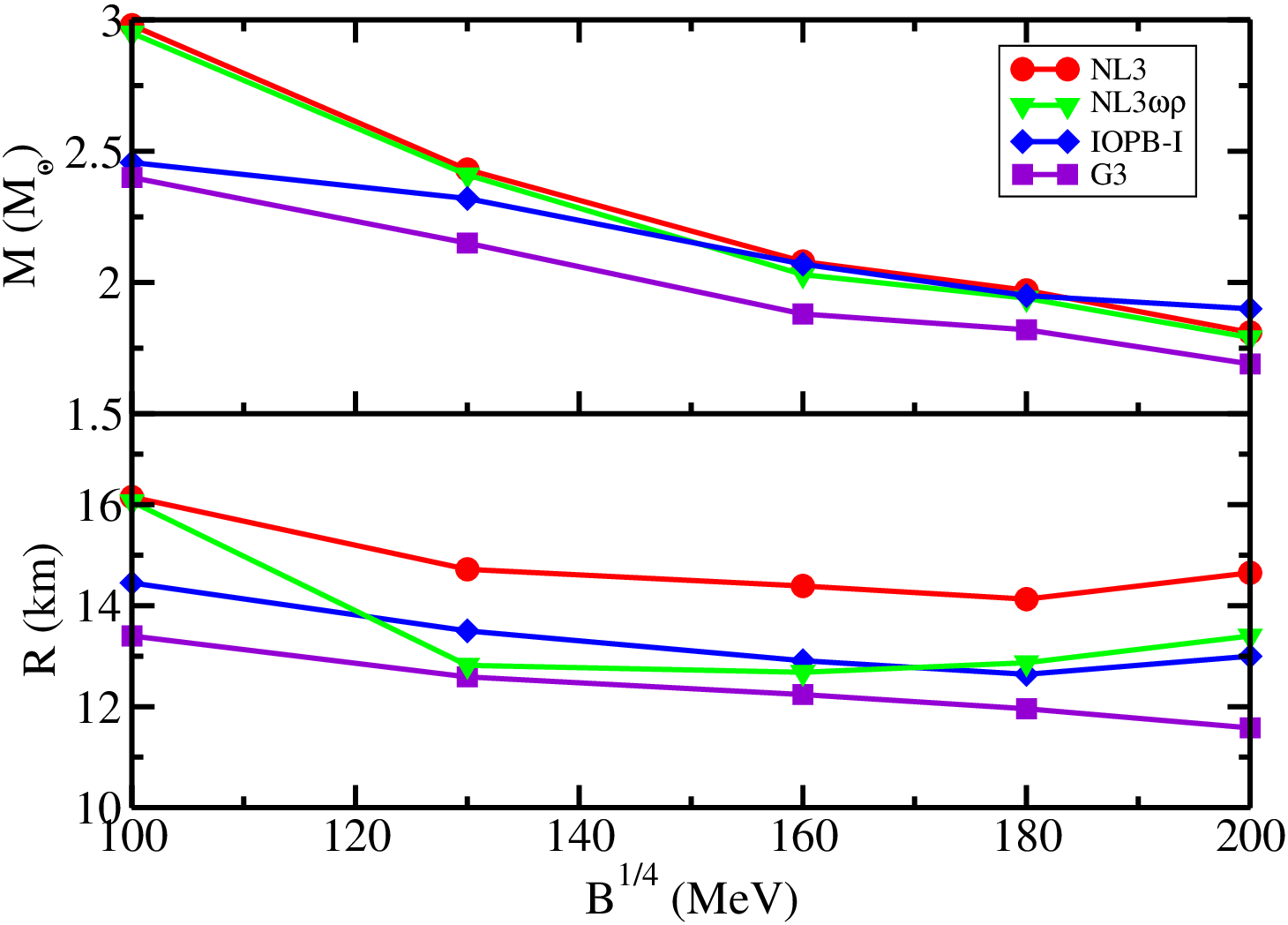}

\end{adjustbox}

 Fig.\ref{mrb} shows the variation of maximum mass and radius with bag constants. The maximum mass for NL3 set at bag value $B^{1/4}$= 100 MeV is 2.98 $M_{\odot}$ while for pure hadronic matter, it predicts a mass of 2.81 $M_{\odot}$. As the bag constant increases, the maximum mass decreases from 2.98 to 1.81$M_{\odot}$. The hybrid NS maximum mass for the NL3${\omega \rho}$ set is almost similar to the NL3 set. However, it predicts a hybrid star with smaller radius as compared to NL3. So, while the  GW170817 data rules out the pure NL3 and NL3${\omega \rho}$ EoSs, the addition of quarks softens the EoS and hence reduces the maximum mass satisfying the GW170817 mass constraints. Similarly, the maximum mass of IOPB-I and G3 set decreases from 2.46 to 1.90$M_{\odot}$ and 2.40 to 1.69 $M_{\odot}$ respectively. From fig. \ref{mrb},we see that for NL3 set with B$^{1/4}$=100 MeV the maximum mass obtained is 2.98M$_{\odot}$ which doesn't agree with the recent mass measurements from gravitational wave data. For IOPB-I and G3 sets, the maximum mass produced at this bag value is 2.46 and 2.40M$_{\odot}$ respectively, which are also larger than the recent mass measurements. Furthermore, for B$^{1/4}$=100 MeV, the transition from hadron matter to quark matter takes place well below the normal nuclear density, predicting the presence of quarks below nuclear density,  which is unphysical. So we agree that the bag value of 100 MeV produces unphysical results and hence isn't consider as a proper choice for the bag constant. The similar results follow for bag value of 200 MeV where the maximum mass produced is much less than 2M$_{\odot}$ and thus this bag values isn't also considered  as a good choice to explain the quark matter in neutron stars.\\
 
   From the gravitational wave observation of the maximum mass of NS in the range 2.01$\pm$0.04 $\le$ $M(M_{\odot})$ $\le$2.16$\pm$0.03, one can see that the bag constant $B^{1/4}$ = 130 MeV produces a maximum mass of 2.15$M_{\odot}$ for G3 set, while the same bag constant gives 2.43$M_{\odot}$ for NL3 set and 2.32$M_{\odot}$ for IOPB-I set. For $B^{1/4}$ = 160 MeV, the maximum mass is 2.08, 2.07 and 1.82 $M_{\odot}$ for NL3, IOPB-I and G3 respectively. The radius of the canonical mass for the bag constants  $B^{1/4}$ = 130-160 MeV is found to be in the range 12.5-13.2 km which is well withing the range of $R_{1.4} \le$13.76 km as extracted from the neutron star tidal deformability of GW170817 event \cite{PhysRevLett.120.172702,PhysRevC.98.035804} . Further, the recently gravitational-wave detected GW190425 constrains the NS mass in the range 1.12 to 2.52 $M_{\odot}$ \cite{collaboration2020gw190425} . Thus we see that the most probable values of bag constant for the obtained EoSs lies in the range 130 MeV$<$$B^{1/4}$$<$ 160 MeV. This range of bag constants determines the properties of hybrid stars that agree with the recent gravitational-wave observations GW170817 and GW190425. A. Aziz $et~al$ \cite{Aziz:2019rgf} has constrained the value of bag constant in the range 150 MeV$\le$$B^{1/4}$$\le$ 180 MeV. The precisely measured mass of 1.97$\pm$0.04 $M_{\odot}$ for binary millisecond PSR J1614-2230 by Demorest $et~al$ \cite{Demorest} supports the presence of quarks in the NS \cite{_zel_2010, Alford,PhysRevC.89.015806,Lastowiecki2015,PhysRevC.87.025804} .
 
\section{Summary and Conclusion}
We studied the hybrid EoS by mixing hadron matter and quark matter using Gibbs conditions. We employed the E-RMF model for hadron matter with recently reported parameter sets and MIT bag model for quark matter with different bag constants. The BPS is used for the outer crust of neutron star. For the inner crust, different EoSs have been used depending upon the model and its nuclear matter properties at saturation. The star matter properties like mass and radius are calculated for different bag constants. We found that the value of bag constant in the range 130 MeV$\le$$B^{1/4}$$\le$ 160 MeV is suitable for explaining the quark matter in neutron stars. The results obtained with bag value less than 130 MeV and greater than 160 MeV do not agree with the recently measured observables from gravitational wave data. Hence such bag values are not suitable enough to explain the quark matter in neutron stars. The increase in the bag value softens the EoS and hence reduces the maximum mass of the star. Other important quantities like tidal deformability will also be affected by the presence of quarks and hence the study of quark matter in neutron stars is of great importance to constrain the EoS. \\

Since the bag constant has a huge effect on the EoS, a more dominant theoretical approach is required to constrain the value of bag constant. The presence of exotic phases like kaons, hyperons etc in the neutron stars will further modify the EoS. The presence of quarks will provide a new insight into the physics of neutron stars and other high dense objects.

\section*{Acknowledgment}
Ishfaq A. Rather is thankful to the Institute of Physics, Bhubaneswar for providing the hospitality during the work. We are thankful to the referees for enhancing our understanding on the subject and the quality of the paper. 

\bibliographystyle{ws-ijmpe}

\bibliography{references}

\end{document}